\documentclass[manuscript,noblind]{geophysics}

\usepackage{multirow}
\usepackage{amsmath}

\usepackage[colorlinks,
linkcolor=blue,
urlcolor = blue,
citecolor=blue
]{hyperref}

\begin{document}

\title{Adaptive graded denoising of seismic data Based on noise estimation and local similarity}

\renewcommand{\thefootnote}{\fnsymbol{footnote}} 


\address{
    \footnotemark[1]Key Laboratory of Earth Exploration and Information Technology of Ministry of Education, Chengdu University of Technology, China. E-mail: 2022020424@cdut.edu.cn\\
    \footnotemark[2]State Key Lab of Oil and Gas Reservoir Geology and Exploitation; School of Geophysics, Chengdu University of Technology, China. E-mail: liyong07@cdut.edu.cn(corresponding author); zhangquanliao@outlook.com; yingtianliu06@outlook.cn; 351587229@qq.com;\\
}

\author{Yang Xueting\footnotemark[1], Li Yong\footnotemark[2], Liao Zhangquan\footnotemark[2], Liu Yingtian\footnotemark[2], Peng Junheng\footnotemark[2].}

\lefthead{Yang et al.} 
\righthead{The graded denoising} 

\maketitle

\begin{abstract}
Seismic data denoising is an important part of seismic data processing, which directly relate to the follow-up processing of seismic data. In terms of this issue, many authors proposed many methods based on rank reduction, sparse transformation, domain transformation, and deep learning. However, when the seismic data is noisy, complex and uneven, these methods often lead to over-denoising or under-denoising. To solve this problems, we proposed a novel method called noise level estimation and similarity segmentation for graded denoising. Specifically, we first assessed the average noise level of the entire seismic data and denoised it using block matching and three-dimensional filtering (BM3D) methods. Then, the denoised data is contrasted with the residual using local similarity, pinpointing regions where noise levels deviate significantly from the average. The remaining data is retained intact. These areas are then re-evaluated and denoised. Finally, we integrated the data retained after the first denoising with the re-denoising data to get a complete and cleaner data. This method is verified on theoretical model and actual seismic data. The experimental results show that this method has a good effect on seismic data with uneven noise.
\end{abstract}

\section{Introduction}

In order to attenuate random seismic noise, most methods are based on various assumptions that reflect or approximate reflect real physics. For example, by assuming that a combination of mathematical basis correspondence lines can fit a physical signal, seismic data can be transformed from a time-domain to another domain, and then the random noise can be attenuated by using the difference between the signal and the random noise in that domain. Such a method is called the domain transformation method. Wavelet transform \citep{stockwell1996localization}, curved wave transform \citep{ma2010curvelet} and other domain transform methods are also the basis of many seismic data noise reduction methods. empirical mode decomposition (EMD) can also be regarded as a generalized domain transformation method, which decomposes the signal according to the time scale characteristics of the signal itself, without setting any mathematical basis function in advance \citep{huang1998empirical}. Based on the assumption of domain transformation, f-x (frequency spatial domain) deconvolution and its development further assume that seismic data is the superposition of linear in-phase axes of different inclination angles, and then carry out noise attenuation by predicting and filtering the seismic in-phase axes in the frequency spatial domain \citep{chen2014random}. Singular Value decomposition (SVD) noise reduction methods, such as multi-channel singular spectrum analysis (f-x MSSA), are the development of matrix decomposition methods in the field of seismic noise removal. Based on the assumption that seismic signals (in phase axes) are always well correlated, it is assumed that seismic signals are concentrated on eigenvectors corresponding to larger eigenvalues \citep{oropeza2011simultaneous}. Methods based on singular value decomposition are also known as low-rank approximation algorithms, and are often combined with domain transform methods such as time-frequency transform and frequence-space transform \citep{nazari2016sparse}. Dictionary learning noise reduction algorithm is a noise reduction method based on sparse representation of data. It assumes that most of the small coefficients in the sparse representation vector of noisy data can be regarded as noise \citep{donoho2006compressed}. Therefore, by removing these noise coefficients, seismic signals can be denoised \citep{yu2015interpolation}. Local feature statistical methods, including median filtering methods \citep{bednar1983applications}, assume that seismic signals are mathematically analyzable within the local window. Among them, the non-local mean algorithm avoids the restriction of local Windows by assuming that the target signal is the weighted average of similar neighborhood structures \citep{bonar2012denoising}.

Although various assumptions play a key role in the attenuation of random noise, in the process of noise attenuation, various assumptions often appear fuzzy when dealing with signals that are not included in the hypothesis, which is likely to cause the effective signal to be removed as noise, resulting in signal leakage in the removed noise. For example, when the energy of the random noise is close to or even greater than that of the effective signal, the hypothesis of the low-rank approximation algorithm is difficult to hold, because the eigenvalue of the noise eigenvector is not necessarily smaller than that of the effective signal eigenvalue. At this time, if the low rank approximation algorithm is used for noise reduction, not only the effective signal will leak, but also the noise removal will be incomplete. For another example, seismic signals such as high-inclination seismic in-phase axes do not fully conform to the hypothesis of linear event superposition, and there will be obvious corner signal leakage in the process of f-x deconvolution denoising \citep{chen2015random}. In addition, in addition to inadequate assumptions, improper selection of parameters is also one of the factors leading to signal leakage. For example, in low-rank methods, the choice of rank is a key step. Choosing a higher rank usually protects more details of the seismic event, but at the same time leaves more residual noise in the denoising results. Choosing a lower rank will produce a clearer output, but at the same time will delete more detailed signals, resulting in signal leakage. Therefore, in general, many noise reduction methods are easy to fall into the dilemma of signal leakage and noise residue due to inadequate assumptions and inappropriate parameter selection.

In fact, in order to reduce signal leakage, scholars proposed some repair methods that compensate the leakage signal back to the noise reduction output \citep{chen2015random}. Among them, the signal leakage used for compensation is detected and extracted through local similarity weight \citep{fomel2007local}. However, the simple signal leakage and noise residue only represent the two end cases of a series of denoising cases, and more cases should be in the noise is not completely removed, but also the signal leakage. In this case, considering the repair method, there are two problems in the extraction and compensation of signal leakage. First of all, the local similarity weight cannot accurately measure the distribution and degree of signal leakage, because in addition to extracting the leakage signal, the residual noise in the output is positively correlated with the noise in the noise removal profile and will also be extracted, resulting in the presence of noise components in the extraction results. Secondly, residual noise in the noise reduction output does not disappear due to compensation signal leakage.

In recent years, although deep neural networks have rapidly become an attractive tool in the field of geophysics \citep{bergen2019machine}, deep learning models often require large amounts of labeled data for training. In noise removal tasks, it can be difficult to obtain high-quality annotated data. So we still tried to improve on the traditional denoising methods that do not need to label data.

Amani\citep{amani2017seismic} used the BM3D method to denoise seismic data. The BM3D method can also achieve good results for uneven noise. But the data examples used are relatively simple. When the structure is complex, it may lead to the loss of effective signals, resulting in some structural changes.

The BM3D algorithm is mainly divided into two steps, the first step is the basic estimation, the second step is the final denoising. The basic estimation of the first step is to provide the weight parameters for the final denoising of the second step. The noise level parameter used in the geophysical application of BM3D is single, which can easily lead to the problem of over-noise and under-noise.

In order to overcome this problem, We have improved the first step of the BM3D method, the basic estimate. Different from the previous single noise evaluation value, we give different noise estimates to regions with different noise levels in order to obtain more accurate noise reduction effect.

\section{theory and methodology}
We use adaptive data segmentation to combine local similarity, noise level estimation and BM3D methods to form a new method. Its complete flow is shown in Figure~\ref{fig:flow}.
\subsection{local similarity}
Local similarity\citep{vulfson2024variant,waclawczyk2024local,zhang2024local}calculated the similarity between adjacent pixels or pixel blocks in two images to find out the similarity between them. Therefore, the region of signal leakage can be determined by local similarity to find the similar part between the image after noise removal and the residual error.

Determine the size and shape of the local area.

For each local region, the correlation coefficient of the corresponding data is calculated (based on Pearson's correlation coefficient calculation formula).
\begin{equation}\label{h1}
	 	r=\frac{\sum\limits_{i=1}^{n}{({{x}_{i}}-\bar{x})}({{y}_{i}}-\bar{y})}{\sqrt{\sum\limits_{i=1}^{n}{{{({{x}_{i}}-\bar{x})}^{2}}}}\sqrt{\sum\limits_{i=1}^{n}{{{({{y}_{i}}-\bar{y})}^{2}}}}}
\end{equation}
Where,${{x}_{i}}$and ${{y}_{i}}$ respectively represent the values of the two variables at the $i$ sample point. $\bar{x}$ and $\bar{y}$ represent the mean of the two variables respectively. $r$ representing the number of samples.

Store the calculated local correlation coefficients in a matrix or array for subsequent analysis and visualization.

\subsection{adaptive classification}
Step 1: Convert the local correlation coefficient into a binary matrix according to 0.2.

Step 2: Remove too small areas to increase processing speed.

Step 3: Draw the initial bounding box according to the connected region of the binary matrix.

Step 4: Check whether any bounding boxes are completely covered and merge the bounding boxes.

Step 5: Extract the incomplete bounding boxes and merge the bounding boxes whose overlapping area exceeds 70\%. And remove the small bounding box to increase the processing speed.

Step 6: Then convert the local correlation coefficient into a binary matrix according to -0.2, and repeat the steps 2-5 above. If the resulting bounding box is the entire data, corrode the data first, and then expand the data to get a proper bounding box.

Step 7: Divide new boundary boxes for overlapping areas to ensure the accuracy of evaluation results.

(All parameters are shown in Table~\ref{tbl:table1})

\subsection{improved BM3D}
The BM3D algorithm is divided into two steps, the first step is the basic estimation, and the second step is the final denoising. The algorithm flow is as follows: 

Step 1: Noise level estimation. Through \citet{liu2013single}, we learned a noise level estimation algorithm base on patch. This approach includes the process of selecting low-rank patches without high frequency components from the seismic data. The selection is based on the gradients of the patches and their statistics. Then, the noise level is estimated from the selected patches using principal component analysis.

Step 2: Final noise removal. The image after the basic estimation is then divided into blocks and estimated block by block.

Grouping. Firstly, block matching is performed on the data obtained from the basic estimation, and a new three-dimensional matrix ${{T}^{'}}_{{{S}_{p}}}$ composed of similar blocks is obtained. Up to this step, there are two 3D arrays, one is a 3D matrix ${{T}_{{{S}_{p}}}}$ composed of similar blocks in the noise image obtained in the first step, and the other is a 3D array ${{T}^{'}}_{{{S}_{p}}}$ composed of similar blocks in the image generated by the basic estimate.

Filtering and denoising. Firstly, the two 3D matrices are transformed simultaneously, then the weights obtained by the basic estimation are filtered and denoised, and then the estimates of all image blocks are obtained by inverse transformation. Finally, the estimates of each pixel are estimated and returned to the original position of the pixel.

The overlapping blocks are revalued and the final result ${{\mu }^{\text{final}}}$ is obtained by weighted average. The expression ${{\mu }^{\text{final}}}$ is as follows:
\begin{equation}\label{h2}
	 	{{\mu }^{final}}(x)=\frac{\sum\limits_{p}{\omega _{P}^{final}}\sum\limits_{Q\in {{S}_{p}}}{{{\chi }_{Q}}}(x)f_{P}^{Q}(X)}{\sum\limits_{p}{\omega _{P}^{final}}\sum\limits_{Q\in {{S}_{p}}}{{{\chi }_{Q}}}(x)},\forall x\in I
\end{equation}

\section{Experimental result} 
In order to prove the feasibility, effectiveness and denoising effect of our method, we denoised the synthetic data with noise generated by theoretical model and the actual seismic data. The model is marmousi model and the actual seismic data is open source actual seismic data.

The signal-to-noise ratio is used as a reference for the denoising effect in the graded denoising experiment\citep{liang2023reinforcement,wu2022seismic}:
\begin{equation}\label{h3}
\text{SNR}(\text{dB})=10\text{lo}{{\text{g}}_{10}}\left( \frac{\|u\|_{2}^{2}}{\left\| \hat{u}-u \right\|_{2}^{2}} \right)
\end{equation}

\subsection{synthetic example} 
In order to test the denoising effect of our method on seismic data with complex structure and uneven noise, we used mamoursi model to generate a seismic data with complex structure. And we also added Gaussian noise with different noise levels to different regions.

Figure~\ref{fig:M_clean} shows the mamoursi model. The parameters of mamoursi model are 13601 channels, sampling interval is 1 ms, and sampling number is 2800. In order to simulate the situation of uneven noise, we added different levels of noise in different regions. As shown in Figure~\ref{fig:M_noise}, we added Gaussian noise with a mean of 0 and a standard deviation of 0.2 to the high noise area, Gaussian noise with a mean of 0 and a standard deviation of 0.05 is added to the low noise area,and Gaussian noise with a mean of 0 and a standard deviation of 0.1 is added to the other regions. After adding noise, the SNR=-0.377.

(a) in Figure~\ref{fig:M_similarity} is the profile obtained after processing by weak texture-based noise level estimation method and BM3D denoising method. It can be seen that in order to be able to self-adapt and accurately divide the regions with large noise levels, we compared the local similarity between the first denoised data and its residuals to get (c) in Figure~\ref{fig:M_similarity}. As shown in (c) in Figure~\ref{fig:M_similarity}, the local correlation coefficient ranges from -1 to 1, so we seted the noise level of region 1 with the local correlation coefficient between -0.2 and 0.2 to be little different from the average noise level of the whole data, and no need to re-denoise. Conversely, region 2 with local correlation coefficient greater than 0.2 and region 3 with local correlation coefficient less than -0.2 are regions where the noise level differs greatly from the average noise level of the data. Our method automatically identify, re-evaluate and denoise areas 2 and 3 as shown in adaptive classification. The process of Figure~\ref{fig:M_block} show the automatic recognition process of regions 2 and 3. Table~\ref{tbl:table2} shows the noise levels for Regions 2 and 3 in some large areas, the average noise levels for the entire data, and the added noise levels. It can be seen from Table~\ref{tbl:table2} that the noise level evaluated after adaptive blocking is closer to the noise level we added. (Both regions 2 and 3 are made up of several small blocks, each with different noise levels.The patches are sorted in order of area from largest to smallest in Table~\ref{tbl:table2})

As shown in Figure~\ref{fig:M_compare}, in the case of complex seismic profile structure and uneven noise, the denoising effect of BM3D method and adaptive median filter is not particularly ideal, and the denoising effect of fx-deconv method and our method is good. But it can be seen from Table ~\ref{tbl:table3} that the signal-to-noise ratio improvement of our method higher than that of fx-deconv method.

\section{Field example} 
As can be seen from Figure~\ref{fig:nd}, the structure in the seismic profile is complex, and the noise is strong and uneven. Some random noise energy exceeds the effective signal, and some geological structures are clearer.

Figure~\ref{fig:nd_indicate} is the seismic profile obtained by the noise level estimation method based on the BM3D denoising method. The random noise in the seismic profile after denoising is basically removed, which makes the entire profile clearer, enhances the structural prominence and layer continuity, and greatly improves the quality of the seismic profile. But where the red box is circled is obviously very blurred, and it is difficult to see the texture. The reason is that the noise level in this part of the region is much lower than the other. However, because the input noise level parameter in BM3D is single, and the source of this parameter is the average noise of the entire seismic data, this area is excessively denoised.(The process of local similarity and adaptive segmentation is shown in Figures~\ref{fig:nd_similarity} and~\ref{fig:nd_block}.)

Table~\ref{tbl:table4} shows the difference between the noise levels of the small patches segmented and the noise levels of the whole data (The patches are sorted in order of area from largest to smallest). Figure~\ref{fig:nd_compare} shows the comparison of our method, adaptive median filtering method, fx-deconv and BM3D method for denoising. It is clear that our method and the resolution of the BM3D method are higher. Moreover, for the area circled by the red box in Figure~\ref{fig:nd_compare}, that is, the patch 3, our method has a better denoising effect. As can be seen from Table~\ref{tbl:table4}, its noise level is much smaller than the noise level of the whole data. Using our method to denoise hierarchically can make the construction clearer and retain more details.

As shown in figure~\ref{fig:Data}, to further validate the effectiveness of our approach, in addition to the open source actual data above, we also applied the following actual data.

It can be seen from Figure~\ref{fig:D_compare} that our method and BM3D method have better denoising effect. Moreover, from the area circled by the black box in Figure~\ref{fig:D_compare} and Figure~\ref{fig:D_patch}, it can be seen that our method has better denoising effect for areas with large noise level difference. (The process of local similarity and adaptive segmentation is shown in Figures~\ref{fig:D_similarity} and~\ref{fig:D_block}. Table~\ref{tbl:table5} shows the difference between the noise levels of the small patches segmented and the noise levels of the whole data. The patches are sorted in order of area from largest to smallest.)


\section{Conclusion}
The noise contained in the actual seismic data is very complex, and the composition and structure of the noise are also complex. Moreover, the noise is not uniform, which leads to the phenomenon of excessive noise removal or residual noise in the smoothing process of conventional denoising methods. The method presented in this paper has certain advantages in dealing with the strong noise with uneven and complex structure. By comparing and analyzing the denoising process of theoretical data and actual data with different denoising methods, the denoising effect of our method is better than that of conventional denoising methods. Moreover, the signal-to-noise ratio of seismic data in many places is very low, which brings great difficulties to the processing and interpretation personnel. The method presented in this paper has a good fidelity to small structures even with strong noise and uneven noise, which lays a good foundation for the interpretation of fine structures in the future.

\section{Data and materials availability}
The original contributions presented in the study are included in the paper; further inquiries can be directed to the corresponding author.

\bibliographystyle{seg}  
\bibliography{graded}

\tabl{table1}{The parameters of adaptive classification}
{
\renewcommand\arraystretch{0.7}
\centering
\begin{center}
\resizebox{0.8\textwidth}{!}{
\begin{tabular}{|c|c|c|c|}
\hline
Binarization threshold & Overlap merge threshold & Area ignore threshold & Boundary box removal threshold \\ \hline
0.2(-0.2)              & 70\%                    & S/24000               & S/720                          \\ \hline
\end{tabular}}
\end{center}
}

\tabl{table2}{The noise level}
{
\renewcommand\arraystretch{0.7}
\centering
\begin{center}
\begin{tabular}{|c|c|c|}
\hline
           & Noise level of estimation & The added noise level         \\ \hline
Whole data & 0.1161                    & A mixture of 0.05,0.1 and 0.2 \\ \hline
Patch 1    & 0.1424                    & A mixture of 0.05,0.1 and 0.2 \\ \hline
Patch 2    & 0.1999                    & 0.2                           \\ \hline
Patch 3    & 0.0998                    & 0.1                           \\ \hline
Patch 4    & 0.0613                    & A mixture of 0.05 and 0.1     \\ \hline
Patch 5    & 0.1000                    & 0.1                           \\ \hline
Patch 6    & 0.0511                    & 0.05                          \\ \hline
Patch 7    & 0.1000                    & 0.1                           \\ \hline
Patch 8    & 0.1038                    & 0.1                           \\ \hline
Patch 9    & 0.0999                    & 0.1                           \\ \hline
Patch 10   & 0.0998                    & 0.1                           \\ \hline
\end{tabular}
\end{center}
}

\tabl{table3}{SNR}
{
\renewcommand\arraystretch{0.7}
\centering
\begin{center}
\resizebox{0.8\textwidth}{!}{
\begin{tabular}{|c|c|c|}
\hline
\multirow{2}{*}{Methods}       & \multirow{2}{*}{\begin{tabular}[c]{@{}c@{}}SNR before denoising\\ SNR/db\end{tabular}} & \multirow{2}{*}{\begin{tabular}[c]{@{}c@{}}SNR after denoising\\ SNR/db\end{tabular}} \\
                               &                                                                                        &                                                                                       \\ \hline
Self-adapting median filtering & -0.377                                                                                 & 3.8671                                                                                \\ \hline
BM3D                           & -0.377                                                                                 & 7.4075                                                                                \\ \hline
Fx-deconv                      & -0.377                                                                                 & 16.8010                                                                               \\ \hline
Our method                     & -0.377                                                                                 & 18.8076                                                                               \\ \hline
\end{tabular}}
\end{center}
}

\tabl{table4}{Noise level}
{
\renewcommand\arraystretch{0.7}
\centering
\begin{center}
\resizebox{0.8\textwidth}{!}{
\begin{tabular}{|c|c|}
\hline
Data       & Noise level of estimation \\ \hline
Whole data & 0.0373                    \\ \hline
Patch 1    & 0.0446                    \\ \hline
Patch 1    & 0.0223                    \\ \hline
Patch 1    & 0.0202                    \\ \hline
Patch 1    & 0.0212                    \\ \hline
Patch 1    & 0.0280                    \\ \hline
\end{tabular}}
\end{center}
}

\tabl{table5}{Noise level}
{
\renewcommand\arraystretch{0.7}
\centering
\begin{center}
\resizebox{0.8\textwidth}{!}{
\begin{tabular}{|c|c|}
\hline
Data       & Noise level of estimation \\ \hline
Whole data & 0.0345                    \\ \hline
Patch 1    & 0.0296                    \\ \hline
Patch 2    & 0.0362                    \\ \hline
Patch 3    & 0.0316                    \\ \hline
Patch 4    & 0.0511                    \\ \hline
Patch 5    & 0.0321                    \\ \hline
Patch 6    & 0.0380                    \\ \hline
Patch 7    & 0.0383                    \\ \hline
Patch 8    & 0.0384                    \\ \hline
Patch 9    & 0.0251                    \\ \hline
Patch 10   & 0.0421                    \\ \hline
\end{tabular}}
\end{center}
}

\newpage
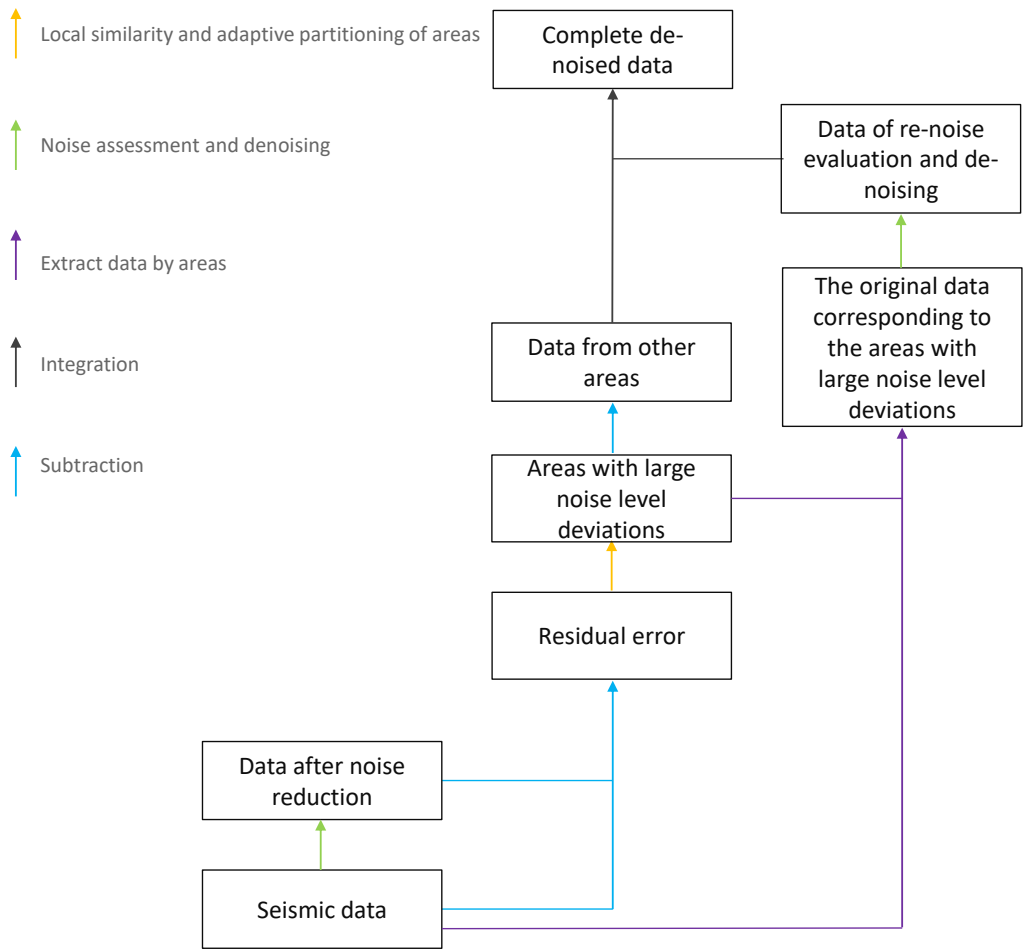
\plot{flow}{width=\textwidth}{The flow of denoising}

\newpage
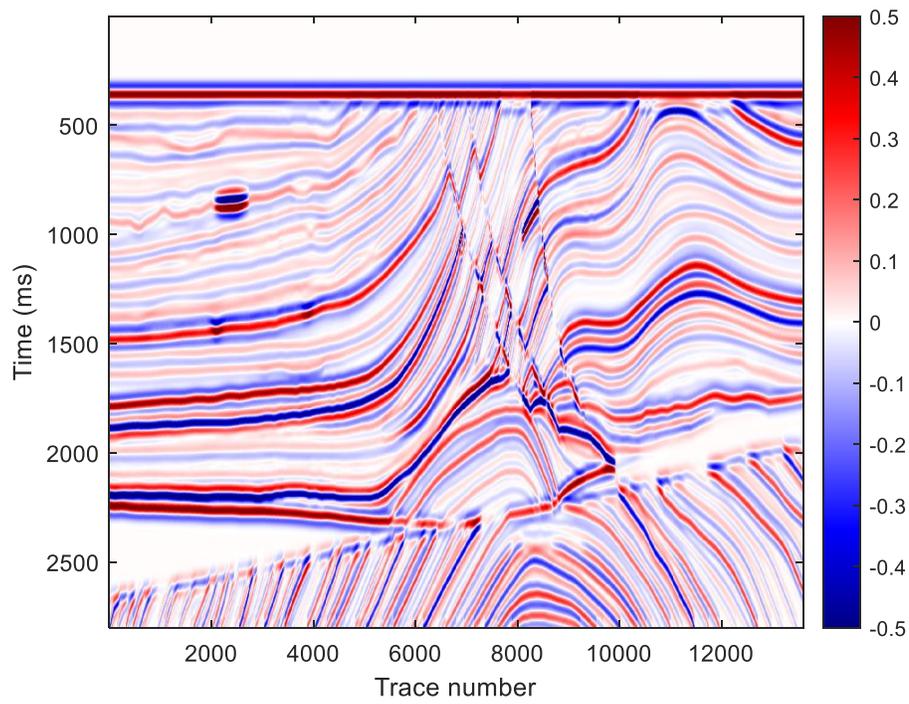
\plot{M_clean}{width=\textwidth}{Marmousi theoretical model}

\newpage
\plot{M_noise}{width=\textwidth}{Noise diagram}

\newpage
\plot{M_similarity}{width=\textwidth}{ Local similarity.(a) Image after first denoising. (b) Residual error. (c) Local similarity}

\newpage
\plot{M_block}{width=\textwidth}{The process of adaptive identification of regions with large noise level differences. (a) to (d) and (e) to (h) represent the process by which our method progressively identifies the positive and negative correlations in local similarity that are worth extracting, respectively. (i) All regions with large noise level differences.}

\plot{M_compare}{width=\textwidth}{Comparison of the denoising results of different algorithms and the residual error. (a), (b), (c) and (d) respectively represent the resultant images after denoising using adaptive median filtering method, the BM3D method, fx-deconv method, and our method. (e), (f), (g) and (h) respectively represent the residuals using adaptive median filtering method, the BM3D method, fx-deconv method, and our method.}

\newpage
\plot{nd}{width=1\textwidth}{An open source actual data}

\newpage
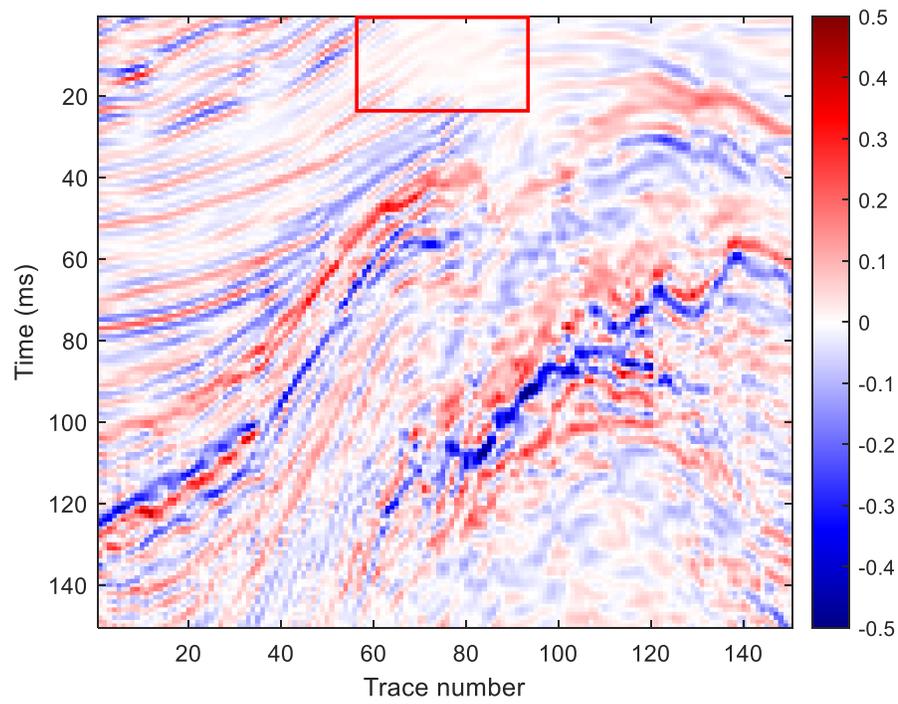
\plot{nd_indicate}{width=1\textwidth}{BM3D denoising result}

\newpage
\plot{nd_similarity}{width=\textwidth}{ Local similarity.(a) Image after first denoising. (b) Residual error. (c) Local similarity}

\newpage
\plot{nd_block}{width=\textwidth}{The process of adaptive identification of regions with large noise level differences.(a) to (d) and (e) to (h) represent the process by which our method progressively identifies the positive and negative correlations in local similarity that are worth extracting, respectively. (i) shows all regions with large noise level differences.}

\newpage
\plot{nd_compare}{width=\textwidth}{Comparison of the denoising results of different algorithms and the residual error. (a), (b), (c) and (d) respectively represent the resultant images after denoising using adaptive median filtering method, the BM3D method, fx-deconv method, and our method. (e), (f), (g) and (h) respectively represent the residuals using adaptive median filtering method, the BM3D method, fx-deconv method, and our method.}

\newpage
\plot{Data}{width=0.9\textwidth}{An open source actual data}

\newpage
\plot{D_similarity}{width=0.7\textwidth}{Local similarity}

\newpage
\plot{D_block}{width=\textwidth}{The process of adaptive identification of regions with large noise level differences.(a) to (d) and (e) to (h) represent the process by which our method progressively identifies the positive and negative correlations in local similarity that are worth extracting, respectively. (i) shows all regions with large noise level differences.}

\newpage
\plot{D_compare}{width=\textwidth}{Comparison of the denoising results of different algorithms and the residual error. (a), (b), (c) and (d) respectively represent the resultant images after denoising using adaptive median filtering method, the BM3D method, fx-deconv method, and our method. (e), (f), (g) and (h) respectively represent the residuals using adaptive median filtering method, the BM3D method, fx-deconv method, and our method.}

\newpage
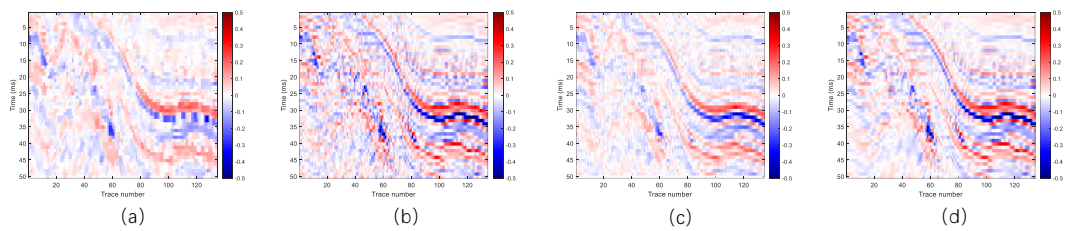
\plot{D_patch}{width=\textwidth}{Patch4 amplification.(a) adaptive median filtering method, (b) the BM3D method, (c)fx-deconv, (d) our method}
\end{document}